\documentclass[twocolumn,a4paper,showpacs,preprintnumbers,amsmath,amssymb]{revtex4-1}

\usepackage{natbib, hyperref}
\usepackage{graphicx}% Include figure files
\usepackage{upgreek}
\usepackage{dcolumn}% Align table columns on decimal point
\usepackage{bm}% bold math
\usepackage{color}

\begin{document}

% Use the \preprint command to place your local institutional report number
% on the title page in preprint mode.
% Multiple \preprint commands are allowed.
%\preprint{}

%Title of paper
\title{Scanning tunneling microscopy with InAs nanowire tips}

\affiliation{\mbox{II. Institute of Physics B and JARA-FIT, RWTH Aachen University, 52056 Aachen, Germany}}
\affiliation{\mbox{Peter-Gr\"{u}nberg Institute (PGI-9) and JARA-FIT, Forschungszentrum J\"{u}lich, 52425 J\"{u}lich, Germany}}

%Authors
\author{Kilian Fl\"{o}hr}\email[corresponding author: ]{floehr@physik.rwth-aachen.de}
\affiliation{\mbox{II. Institute of Physics B and JARA-FIT, RWTH Aachen University, 52056 Aachen, Germany}}

\author{Kamil Sladek}
\author{H. Yusuf G\"{u}nel}
\author{Mihail Ion Lepsa}
\author{Hilde Hardtdegen}
\affiliation{\mbox{Peter-Gr\"{u}nberg Institute (PGI-9) and JARA-FIT, Forschungszentrum J\"{u}lich, 52425 J\"{u}lich, Germany}}

\author{Marcus Liebmann}
\affiliation{\mbox{II. Institute of Physics B and JARA-FIT, RWTH Aachen University, 52056 Aachen, Germany}}

\author{Thomas Sch\"{a}pers}
\affiliation{\mbox{Peter-Gr\"{u}nberg Institute (PGI-9) and JARA-FIT, Forschungszentrum J\"{u}lich, 52425 J\"{u}lich, Germany}}

\author{Markus Morgenstern}
\affiliation{\mbox{II. Institute of Physics B and JARA-FIT, RWTH Aachen University, 52056 Aachen, Germany}}

\date{\today}

\begin{abstract}
% insert abstract here
Indium arsenide nanowires grown by selective-area vapor phase epitaxy are used as tips for scanning tunneling microscopy (STM). The STM tips are realized by positioning the wires manually on the corner of a double cleaved gallium arsenide wafer with sub-$\upmu$m precision and contacting them lithographically, which is fully compatible with further integrated circuitry on the GaAs wafer.
STM images show a $z$ noise of 2\,pm and a lateral stability of, at least, 0.5\,nm on a Au(111) surface. $I(z)$ spectroscopy reveals an exponential decay indicating tunneling through vacuum. Subsequent electron microscopy images of the tip demonstrate that the wires are barely modified during the STM imaging. 
\end{abstract}

\pacs{}% insert suggested PACS numbers in braces on next line

\maketitle %\maketitle must follow title, authors, abstract and \pacs
%\thispagestyle{headings}  % Adding page number to the title page
% Body of paper goes here.

%\section{Introduction}
Scanning tunneling microscopy (STM) is the most advanced tool to probe the local density of states \cite{Eigler, Manoharan, Hashimoto}, excitations \cite{Ho, Heinrich} including dynamics \cite{Loth} as well as fingerprints of correlations \cite{Berndt, Crommie, Davis,Yazdani,Yazdani2} down to the atomic scale. However, since the tip is typically metallic,  all long-range correlations are effectively screened cutting, e.g., the electron-electron interaction in quantum Hall systems at a length scale of about \mbox{10\,nm}.\cite{Becker} Using materials with a charge density lower than the tunneling tip (e.g. doped semiconductors) would partially resolve this problem and, in addition, might open the possibility of integrated circuitry directly at the tip. However, little effort has been devoted to semiconducting tips so far.\cite{Wong2012} One appealing possibility is to use the well-developed epitaxially grown semiconductor nanowires.\cite{Law,Poole,Hardtdegen} Such nanowires have already been used for advanced electronic applications and fundamental experiments in quantum physics.\cite{Samuelson2,Ensslin,Lieber,Kouwenhoven}\\
However, several challenges have to be overcome to reach this goal. Especially the preconditions of the same quality in mechanical stability, electrical characteristics and the possibility of atomic resolution as conventional metallic tips have to be achieved. To this end, we show that single InAs nanowires, known for their conducting surfaces without Schottky barriers,\cite{Aristov} are suitable for STM tips. Atomic resolution images of Au(111) and a z-noise of 2\,pm demonstrate the high mechanical stability of these tips. Spectroscopic characterization of the tips reveal the characteristic fingerprints of InAs surfaces.\cite{JESRP}\\

\begin{figure}
\includegraphics[width=8.5cm]{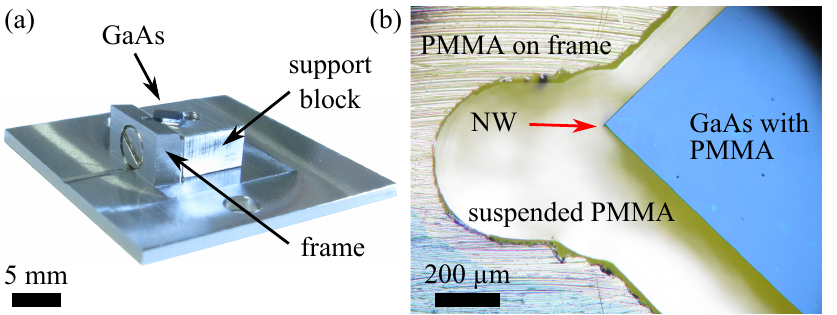}% Here is how to import EPS art, EPS: Latex-ps-pdf
\caption{\label{fig1}(Color online) (a) Piece of GaAs wafer glued on the support block and mounted inside the protective frame before transferring the polymethylmethacrylate (PMMA) sheet. (b) Optical microscope image of the GaAs support inside the protective frame after transferring the PMMA. This covers the frame, the whole GaAs area and the gap in between including the InAs nanowire. The GaAs appears homogeneously blue which, since caused by interference, indicates that the PMMA film is uniform. The position of the invisible InAs nanowire (NW) is marked.}
\end{figure}

%Sample preparation
For the fabrication of our STM tips, we use InAs nanowires which are grown epitaxially by catalyst-free selective area metal organic vapor phase epitaxy.\cite{Hardtdegen, Sladek2012}
The nanowires are n-doped (\mbox{$n \approx$ 2$\times$10$^{18}$\,cm$^{-3}$}, \mbox{$\mu \approx$ 500\,$\frac{\mbox{cm}^2}{\mbox{Vs}}$}) as determined by four-terminal transport measurements, have diameters of approximately 120\,nm and are about 3\,$\upmu$m long. Due to the growth mechanism, the wires consist of frequently alternating wurtzite and zinc blende structure, which, however, have very similar electronic properties.\cite{Zanolli, Frielinghaus, Wirths2011}
As a support for the InAs nanowires we use a semi-insulating GaAs(100) wafer, which is double cleaved in the $<$110$>$ perpendicular directions. The cleavage leads to a sharp edge inside a prepatterned marker field. For better handling, the cleaved piece is glued to a stainless steel block, which can be mounted either on a standard STM tip holder or in a protective frame as shown in Fig.\ \ref{fig1}(a). 
We use a sharp indium tip, mounted onto a micrometer screw to place the nanowire at the desired position. Under an optical microscope a single nanowire is firstly removed from the growth substrate by its adhesive force and subsequently placed at the corner of the GaAs support with a precision of less than 1\,$\upmu$m. This method is described in detail elsewhere.\cite{Floehr2011} We place the wire such that it is freely suspended for approximately 1\,$\upmu$m beyond the corner. This is long enough to guarantee that the wire protrudes the GaAs support when approaching a surface within the STM, but short enough to guarantee mechanical stability of the wire.

\begin{figure}
\includegraphics[width=8.5cm]{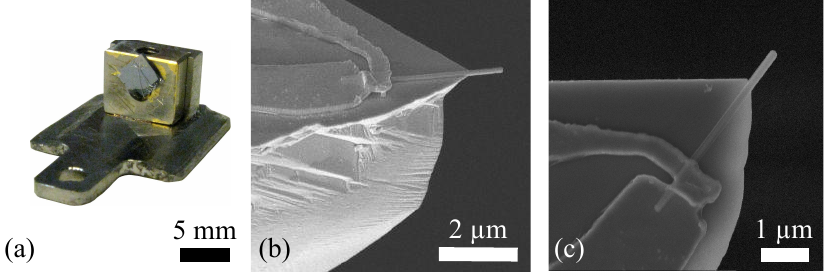}% Here is how to import EPS art, EPS: Latex-ps-pdf
\caption{\label{fig2}(Color online) (a) GaAs support with contacted InAs tip mounted on the STM tip holder. (b) SEM image (75$^\circ$-tilted view) of the STM tip before approaching a Au(111) sample. The small overhang of the GaAs support corner provides a tolerance of about $2.5^\circ$ in angular adjustment accuracy of the nanowire with respect to the probed surface, still guaranteeing a protruding nanowire during the tip approach. (c) SEM image (top view) of the same nanowire after use as a STM tip on a Au(111) surface.}
\end{figure}

The next challenge is to realize ohmic contacts to the nanowire since common spin coating for subsequent lithography results in an inhomogeneous thickness of the polymethylmethacrylate (PMMA) at the sample edges. Different \mbox{PMMA-free} techniques have been tested, in particular indium microsoldering\cite{Girit2007,Geringer2010} and electron beam induced deposition (EBID)\cite{Randolph2006, Bauerdick2006}. Both, however, led to poor electrical contacts with resistances larger than $1$\,M$\Omega$. This is probably due to the native oxide surrounding the InAs nanowire surface.\cite{Oxide1, Oxide2} Therefore, we adopted the technique of Dean \textit{et al.}\ used to transfer graphene onto a boron nitride flake with the aid of a PMMA sheet:\cite{Dean} A polyvinyl alcohol (PVA) 4\,\% solution is spin coated with 4000\,rpm to a (15x15)\,mm$^2$ Si/SiO$_2$ wafer piece and baked for 2\,min at 100$^\circ$C on a hot plate. Subsequently, PMMA (600k 7\,\%) is spin coated with 7000\,rpm on the PVA layer and baked for 7\,min at 120$^\circ$C. The thicker edges of the PMMA are removed with a scalpel and the wafer with the two coatings is brought onto the surface of deionized water to separate the PMMA layer from the substrate by dissolving the PVA. Afterwards, the floating PMMA is fished from the water surface using a metal frame. After drying, the PMMA film is finally transferred mechanically on top of the GaAs support, which is mounted inside the protective frame [Fig.\ \ref{fig1}(a)], and heated to approximately 130$^\circ$C during transfer.
This process results in a largely homogeneous layer of PMMA covering the whole GaAs including the edges and especially the suspended part of the nanowire, which is additionally  stabilized by the continuous film [Fig.\ \ref{fig1}(b)].\\
Afterwards, two contact areas at the end of the supported nanowire are defined by electron beam lithography. After Ar sputtering (0.3\,keV) to remove the native oxide of the wire, contacts are deposited by evaporation of Ti/Au (10\,nm/170\,nm). Nearly all prepared wires survived the subsequent lift-off process using acetone (approx.\ 8 hours) and isopropanol. Figure \ref{fig2}(b,c) shows scanning electron microscopy (SEM) images of such a contacted nanowire before and after usage as an STM tip. Unfortunately the two contacts which were meant to test a sufficient conductivity overlap for this tip which was used for the measurements here. But tests with other nanowires revealed reproducible ohmic behavior of identically prepared contacts exhibiting two-terminal resistances below 20\,k$\Omega$. One four-terminal resistance measurement on the GaAs support even revealed a contact resistance below 1\,k$\Omega$. More importantly, Fig. \ref{fig2} demonstrates that the nanowire survived the approach to and removal from the Au(111) surface implying that the STM measurement in between has been performed by the nanowire.

For STM measurements, the support block with the processed InAs tip is mounted on the tip holder of a home-built room temperature STM system operating in ultrahigh vacuum (10$^{-8}$\,Pa). Using the geometry shown in Fig.\ \ref{fig2}(b), we estimate that the InAs nanowire will reach the surface of the sample first, as long as it is oriented perpendicular to the surface within a deviation of less than $2.5^\circ$. Furthermore the approach should be as perpendicular as possible to minimize the lateral bending forces on the nanowire originating from the sample.
Therefore, we adjusted the surface of the GaAs support with an accuracy of $1^\circ$ safely preventing that this will reach the sample surface first.
In order to remove the oxide barrier at the end of the nanowire, Ar sputtering (ion energy: 0.3\,keV) has to be applied in-situ before approaching the tip to the sample. Indeed, we did not achieve stable imaging without this sputtering step. The wires are firstly ion bombarded under two opposite angles of $\pm$45$^\circ$ to the axial direction, for 30\,min each, and afterwards at 0$^\circ$ for about 60\,min. SEM pictures, recorded directly after sputtering, show an abrasion of less than 10\,nm in diameter and length of the wire.

\begin{figure}
\includegraphics[width=8.5cm]{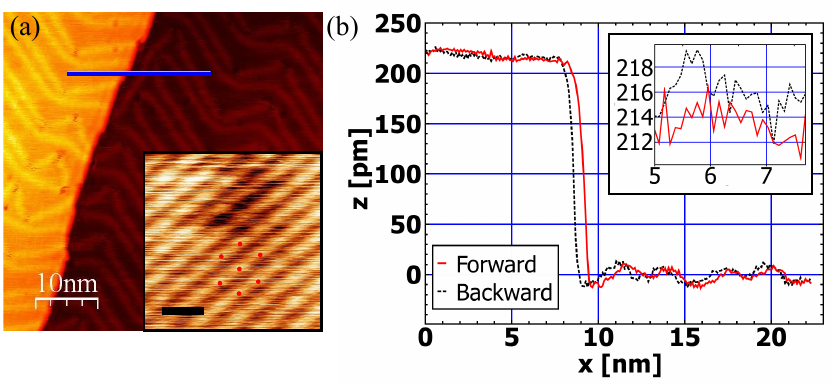}% Here is how to import EPS art, EPS: Latex-ps-pdf
\caption{\label{fig3}(Color online) (a) STM topography of Au(111) (sample bias: \mbox{U = +1\,V}, \mbox{I = 50\,pA}). Inset: STM images exhibiting atomic resolution (\mbox{U = +1\,V}, \mbox{I = 10\,pA}) with red dots marking the atoms. The scale bar is 5\,\AA. (b) Line scan along the blue line in (a). Full red curve: forward scan, dashed black curve: backward scan. The relative shift of the step edge is 0.6\,nm. Inset: zoom into the range from 5\,nm to 8\,nm.}
\end{figure}

\begin{figure}
\includegraphics[width=8.5cm]{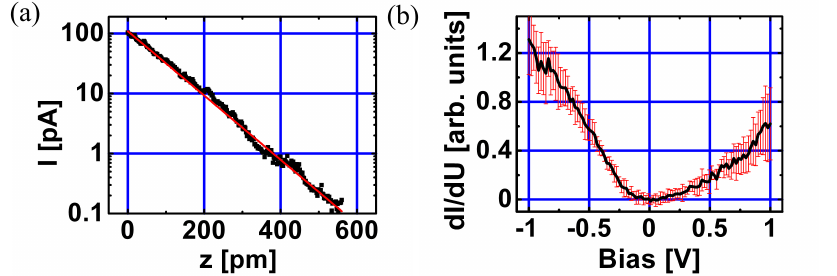}% Here is how to import EPS art, EPS: Latex-ps-pdf
\caption{\label{fig4}(Color online) (a) $I(z)$ curve averaged over 5 spectra recorded with the InAs tip (black squares) at different positions of the Au(111); stabilization at \mbox{U = 1\,V}, \mbox{I = 100\,pA}; an
exponential fit $\mbox{e}^{-2\upkappa \mbox{z}}$ with $\kappa$ = 6.17$\times$10$^9$\,m$^{-1}$ (red line) is shown for comparison. (b) $dI/dU(U)$ curve averaged from 10 spectra;
stabilization at \mbox{I = 100\,pA}, \mbox{U = 1\,V}.}
\end{figure}

%Measurements
STM images are recorded on a Au(111) crystal prepared by several cycles of ion bombardment (600 eV Ar) and annealing ($450^\circ$C). To approach the tip to the Au surface, we use a sample bias of +1\,V, safely avoiding tunneling into the band gap of the nanowire (tunnel current: 10\,pA). Figure \ref{fig3}(a) shows an STM image of a single atomic step of the Au(111) surface. The well-known herringbone reconstruction\cite{Barth} is visible on both terraces. The STM image represents the raw data except that it has been flattened using a plane fit.
The quality of the STM data is clearly seen in the line scan of Fig.\ \ref{fig3}(b) revealing a height difference between the two terraces of (224$\pm$26)\,pm, as determined using histograms of the $z$ values on both terraces. This is in good agreement with the known value of 235\,pm. The width of the step obtained by averaging a few line scans is (0.6$\pm$0.1)\,nm, which can be regarded as an upper limit of the lateral resolution. In particular, it excludes that the nanowire oscillates laterally by an amplitude of more than 0.5\,nm. This is an important finding, rather difficult to achieve, e.g., for carbon nanotube tips.\cite{Dai,Wilson}
Forward and backward line scans exhibit a relative lateral shift of about (0.53$\pm$0.07)\,nm, which is nearly identical to the shift observed by using PtIr tips, (0.65$\pm$0.26)\,nm. The difference is probably due to a remaining creep of the piezo scanner. This excludes, in addition, a strong bending of the nanowire at the step edge or during the scanning process.\\
The noise level in $z$ direction is determined as the standard deviation from the mean value of several line scans recorded on the flat areas of the surface, i.e., in between the reconstruction lines [see inset of Fig.\ \ref{fig3}(b)]. It is \mbox{$\upsigma_z$ = 2.2\,pm} at a remaining current noise of \mbox{$\upsigma_I$ = 1.7\,pA} for a current of I = 50\,pA. 
For a room temperature STM system, this excellent value points again to the high mechanical stability of the nanowire tip.
Atomic resolution on the Au(111) surface has been occasionally observed. One example is shown in the inset of Fig.\ \ref{fig3}(a). The structure did not change by rotating the scan direction and varying the scan speed excluding that it is a measurement artifact. However, the atomic distance is about 5-10\,\% less than the known distance of Au(111), probably due to thermal drift. The apparent line structure of the atoms might be related to a tilted p-orbital at the end of the imaging nanowire.\\
$I(z)$ spectroscopy results are shown in Fig.\ \ref{fig4}(a) revealing an exponential decay of the current with tip-sample separation, as expected for tunneling. The decay constant of \mbox{$\kappa$ = 6.17$\times$10$^9$\,m$^{-1}$} is nearly identical to the one observed for W tips on InAs(110),\cite{JESRP} but slightly reduced with respect to metals probably due to surface band bending effects on the InAs surface. This implies a favorable low charge density at the tip surface. The $dI/dU$ curve shown in Fig. \ref{fig4}(b) is recorded using a lock-in amplifier. It exhibits a reduced $dI/dU$ signal within about 300\,meV around the Fermi level as expected from the InAs band gap.  Importantly, it strongly deviates from spectra expected on GaAs\cite{Feenstra,Feenstra1994} (band gap: 1.42\,eV) additionally evidencing that the InAs nanowire and not the GaAs support is used for STM imaging.

%Conclusion
In summary, we have fabricated InAs nanowire tips for STM measurements exhibiting a rms $z$ noise level of only 2\,pm and a lateral resolution better than 0.5\,nm. $I(z)$ spectroscopy reveals the tunneling characteristics of the tip and subsequent electron microscopy images show its stability during coarse approach and removal from the substrate. All these results show the good quality of STM imaging with a semiconducting nanowire in comparison to well prepared metallic tips at room temperature in ultrahigh vacuum. The InAs tips are placed and lithographically contacted on a GaAs wafer opening the possibility for integrated circuitry directly at the tip.

We thank Mike Pezzotta and Florian Muckel for technical assistance and acknowledge financial support by the excellence initiative of the German federal government.

Copyright 2012 American Institute of Physics. This article may be downloaded for personal use only. Any other use requires prior permission of the author and the American Institute of Physics. The following article appeared in Appl. Phys. Lett. 101, 243101 (2012) and may be found at \url{http://link.aip.org/link/?APL/101/243101}.

%\bibliography{Literaturverzeichnis}

%

\end{document}